\documentclass[preprint]{aastex}
\slugcomment{2016.11 v4}

\shorttitle{Magnetar fallback disk systems} \shortauthors{H. Tong et al.}

\begin{document}

\title{Rotational evolution of magnetars in the presence of a fallback disk}

\author{H. Tong\altaffilmark{1}, W. Wang\altaffilmark{2}, X. W. Liu\altaffilmark{3}, R. X. Xu\altaffilmark{4}}

\altaffiltext{1}{Xinjiang Astronomical Observatory, Chinese Academy of Sciences, Urumqi, Xinjiang 830011,
    China\\ {\it tonghao@xao.ac.cn}}
\altaffiltext{2}{National Astronomical Observatories, Chinese Academy of Sciences, Beijing 100012, China}
\altaffiltext{3}{School of Physics, China West Normal University, Nanchong 637002, China}
\altaffiltext{4}{School of Physics, Peking University, Beijing 100871, China}

\begin{abstract}
Magnetars may have strong surface dipole field. Observationally, two magnetars may have passive fallback disks. In the presence of a fallback disk, the rotational evolution of magnetars may be changed. In the self-similar fallback disk model, it is found that: (1) When the disk mass is significantly smaller than $10^{-6} \,\rm M_{\odot}$, the magnetar is unaffected by the fallback disk and it will be a normal magnetar. (2) When the disk mass is large, but the magnetar's surface dipole field is about or
below $10^{14} \,\rm G$, the magnetar will also be a normal magnetar. A magnetar plus a passive fallback disk system is expected. This may correspond to the observations of magnetars  4U 0142$+$61, and 1E 2259$+$586. (3) When the disk mass is large, and the magnetar's surface dipole field is as high as $4\times 10^{15} \,\rm G$, the magnetar will evolve from the ejector phase to the propeller phase, and then enter rotational equilibrium.  The magnetar will be slowed down quickly in the propeller phase. The final rotational period can be as high $2\times 10^4 \,\rm s$. This may correspond to the super-slow magnetar in the supernova remnant RCW 103. Therefore, the three kinds of magnetars can be understood in a unified way. 
\end{abstract}


\section{Introduction}

Magnetars are supposed to be neutron stars powered by their strong magnetic field (Duncan \& Thompson 1992). Observationally,  they can have X-ray luminosity as high as $10^{35} \,\rm erg \,s^{-1}$, rotational period in the range of $2$ to $12$ seconds. Their period derivative can be as high as $10^{-11} \,\rm s\, s^{-1}$. This may be due to their strong surface dipole field, which may lie in the range of $10^{14}-10^{15} \,\rm G$ (Olausen \& Kaspi 2014). The multipole field of magnetars can be even higher than their surface dipole field. 
The release of magnetic energy powers both the bursts (including giant flares), and persistent emissions of magnetars
(Beloborodov 2009; Vigano et al. 2013).  Previously, two magnetars 4U 0142$+$61 and 1E 2259$+$586 are found to have relatively low surface dipole field compared with other normal magnetars\footnote{The magnetic dipole field of 4U 0142$+$61 is about $1.3 \times 10^{14} \,\rm G$. It is only higher than five other magnetars: 1E 2259$+$586, CXOU J164710.2$-$455216, and the three low magnetic field magnetars (Olausen \& Kaspi 2014). The low magnetic field magnetars may form a separate problem of magnetar researches (Rea et al. 2010). SGR 1806$-$20 has the highest characteristic magnetic field, about $2.5 \times 10^{15} \,\rm G$.}. Incidentally, these two magnetars are also found to have passive fallback disks (Wang et al. 2006; Kaplan et al. 2009). A disk may be formed when some of the supernova ejecta material fall onto the central compact star, i.e. a fallback disk (Perna et al. 2014). If the disk mass is negligibly small, then the central star can be said to have no fallback disk.

Recently, the central compact object (CCO) in supernova remnant RCW 103 is found to be a magnetar (D'Ai et al. 2016; Rea et al. 2016). The magnetar-like burst, outburst, and transient hard X-ray emissions all resemble those of classical magnetars (Rea \& Esposito 2011).  However, this magnetar (1E 161348$-$5055) may have a rotational period of $6.6$ hours (De Luca et al. 2006). The most recent observations also point to a rotational origin for this very long period (D'Ai et al. 2016; Rea et al. 2016). This super-slow magnetar may be spun-down by additional torques from a fallback disk (De Luca et al. 2006; Li 2007; Rea et al. 2016). However, no detailed calculations are available at present\footnote{Li (2007) only present a general Monte Carlo simulation. They did not point out which physical parameters in their simulations result in the very long rotational period.}. Furthermore, it is still unknown how to understand in a unified way the three different kinds of magnetars at present: normal magnetars, magnetars with passive fallback disks, and the super-slow magnetar with a rotational period of $6.6$ hours. 

In this paper, the rotational evolution of magnetars in the presence of a fallback disk is calculated. It is found that the three different kinds of magnetars can by understood together in terms of a different combination of fallback disk mass and magnetic dipole field. 

\section{Rotational evolution of magnetars in the presence of a fallback disk}

\subsection{Description of the fallback disk model}

One aspect of the fallback disk model is to explain the persistent X-ray and timing observations of magnetars.  
In previous fallback disk models (Chatterjee et al. 2000; Alpar 2001; Benli \& Ertan 2016), the central star is a normal neutron star with surface magnetic dipole field of $10^{12}-10^{13} \,\rm G$. The accretion from the fallback disk provides both the X-ray luminosity and spin-down torque of the neutron star. There is no magnetar in these models. Similar description of the fallback disk as that of Chatterjee et al. (2000) is adopted in the following calculations. The major difference is that the normal neutron star is replaced by a magnetar, whose magnetic dipole field can be as high as $10^{15} \,\rm G$. 

The mass accretion rate due to a fallback disk will be a decreasing power law form in the self-similar solution (Cannizzo et al. 1990; Chatterjee et al. 2000). It is constant during an initial short time, and decay in a power law form there after:
\begin{eqnarray}
\dot{M}&=& \dot{M}_0, \quad  \quad \quad 0<t<T, \\ \nonumber
             &=& \dot{M}_0 \left(  \frac{t}{T} \right)^{-\alpha}, \quad  t\ge T,
\end{eqnarray}
where $\dot{M}$ is the mass accretion rate (which is actually the mass transfer rate at the outer edge of the disk), $\dot{M}_0$ is the initial mass accretion rate, $T$ is a typical time scale, $\alpha$ is the power law index, and $t$ is the age. The initial mass accretion rate is related to the initial disk mass
\begin{equation}
M_{\rm d,0} = \frac{\alpha}{\alpha-1} \dot{M}_0 T,
\end{equation}
where $M_{\rm d,0}$ is the initial disk mass. Compared with Chatterjee et al. (2000), four modifications are considered:
\begin{enumerate}
\item The time scale $T$ is set to be the dynamical time scale in Chatterjee et al. (2000), which is about 1 millisecond. 
According to Menou et al. (2001), the time scale may be determined by the viscous time scale
\begin{equation}
T \approx 6.6 \times 10^{-5} \,\rm yr \approx 2000 \,\rm s,
\end{equation}
for typical parameters of the disk. 

\item The power law index is $\alpha=19/16$ for opacity dominated by electron scattering, and $\alpha=1.25$ for a Kramers opacity (Cannizzo et al. 1990). Chatterjee et al. (2000) chose $\alpha=7/6$. While during the main life time of the fallback disk, the Kramers opacity may be more relevant (Francischelli \& Wijers 2002; Li 2007) and $\alpha$ should be $1.25$.

\item During the early time, the mass accretion rate can be highly super-Eddington. Chatterjee et al. (2000) used the mass accretion rate without considering the Eddington limit. 
According to Yan et al. (2012), the mass accrete at the inner edge of the disk should be limited by the Eddington accretion rate, which is chosen as $\dot{M}_{\rm Edd} =10^{18} \,\rm g\, s^{-1}$ in the following calculations. Denote the time when the mass accretion rate equals the Eddington accretion rate as $t_{\rm eq}$, the mass accretion rate at the inner edge of the disk is: 
\begin{eqnarray}\label{Mdotacc}
\dot{M}_{\rm acc}&=& \dot{M}_{\rm Edd}, \quad  \quad \quad \ \ 0<t<t_{\rm eq}, \\ \nonumber
             &=& \dot{M}_{\rm Edd} \left(  \frac{t}{t_{\rm eq}} \right)^{-\alpha}, \quad  t\ge t_{\rm eq}.
\end{eqnarray}
This mass accretion rate $\dot{M}_{\rm acc}$ determines the interaction between the fallback disk and the neutron star. Since $t_{\rm eq}$ is always larger than $T$, it is
\begin{equation}\label{teq}
t_{\rm eq} =T \left(  \frac{\dot{M}_0}{\dot{M}_{\rm Edd}}  \right)^{1/\alpha}. 
\end{equation}
For a chosen time scale $T$, the mass accretion rate $\dot{M}_{\rm acc}$ is determined by the initial disk mass, which may range from 
$10^{-6}\,\rm M_{\odot}$ to $0.1 \,\rm M_{\odot}$ (Michel 1988; Chevalier 1989; Wang et al. 2006; Perna et al. 2014). 

\item In the case of a high magnetic field, the scattering cross section between electrons and photons may be significantly suppressed (Paczynski 1992). The corresponding critical luminosity and accretion rate for an accreting magnetar can be $10^2$ to $10^4$ times higher than the traditional non-magnetic case (Tong 2015; Mushtukov et al. 2015).  However, the following calculations is not affected by a different value of  critical accretion rate. For a higher critical accretion rate, the timescale $t_{\rm eq}$ will be smaller. The mass accretion rate at later time will be the same, as can be seen by substituting eq.(\ref{teq}) to the second line of eq.(\ref{Mdotacc}).  The numerical calculation is also consistent with this analysis. Therefore, this degree of freedom can be neglected. 
    
\end{enumerate}

The rotational evolution of the neutron star is determined by the interaction between the accretion flow and neutron star's magnetosphere
\begin{equation}
I \dot{\Omega} = N,
\end{equation}
where $I$ is the neutron star moment of inertia, set to be $10^{45} \,\rm g \,cm^2$, $\Omega$ is the angular velocity of the neutron star, and $N$ is the torque. 
When the magnetospheric radius\footnote{The radius where the magnetic energy density of the neutron star equals the kinetic energy density of the accretion flow (Shapiro \& Teukolsky 1983; Lai 2014): $R_{\rm m} =(\mu^4/2GM \dot{M}_{\rm acc}^2)^{1/7}$, where $G$ is the gravitational constant, and $M$ is the neutron star mass, set to be $1.4\,\rm M_{\odot}$.} is larger than the light cylinder radius\footnote{The radius where the corotational velocity equals the speed of light: $R_{\rm lc} =P c/2\pi$, where $P$ is the rotational period of the neutron star.}, the neutron star will be unaffected by the fallback disk. 
 In this ejector phase, the magnetar will be a normal magnetar. The torque in this case can be approximated by the magnetic dipole braking 
\begin{equation}
N_{\rm d} = -\frac{2\mu^2 \Omega^3}{3c^3},
\end{equation}
where $\mu$ is the magnetic dipole moment, and $c$ is the speed of light. The magnetic dipole moment is determined by the neutron star surface magnetic dipole field and radius: $\mu =B R^3$. When the magnetospheric radius is smaller than the light cylinder radius but larger than the corotation radius\footnote{The radius where the local Keplerian velocity is equal to the corotational velocity: $R_{\rm co} =(G M/4\pi^2)^{1/3} P^{2/3}$.}, the neutron star is in the propeller phase (Illarionov \& Sunyaev 1975). In the propeller phase, the neutron star will lose angular momentum by pushing away the accreted matter. When the magnetospheric radius is smaller than corotation radius, the accretion flow can fall onto the neutron star. In this accretion phase, the neutron star will gain angular momentum. A unified torque for both the accretion phase and the propeller phase is (Menou et al. 1999; Chatterjee et al. 2000)
\begin{equation}\label{propeller_torque}
N_{\rm prop} = 2\dot{M}_{\rm acc} R_{\rm m}^2 \Omega_{\rm K}(R_{\rm m}) \left(  1- \frac{\Omega}{\Omega_{\rm K}(R_{\rm m})} \right), 
\end{equation}
where $\Omega_{\rm K}(r)=(GM/r^3)^{1/2}$ is the Keplerian angular velocity. In the propeller phase, the spin-down torque is proportitonal to $\propto -\Omega$. This will result in an exponential increase of the rotational period. Therefore, in principle, a very long rotational period of a young neutron star is possible. For a typical initial disk mass of $10^{-5} \,\rm M_{\odot}$, the magnetic field evolution due to accretion is not significant. The magnetic field is assumed to be constant in the following calculations. 

\subsection{Calculations for typical magnetar parameters}

\begin{figure}[htbp]
\centering
\includegraphics{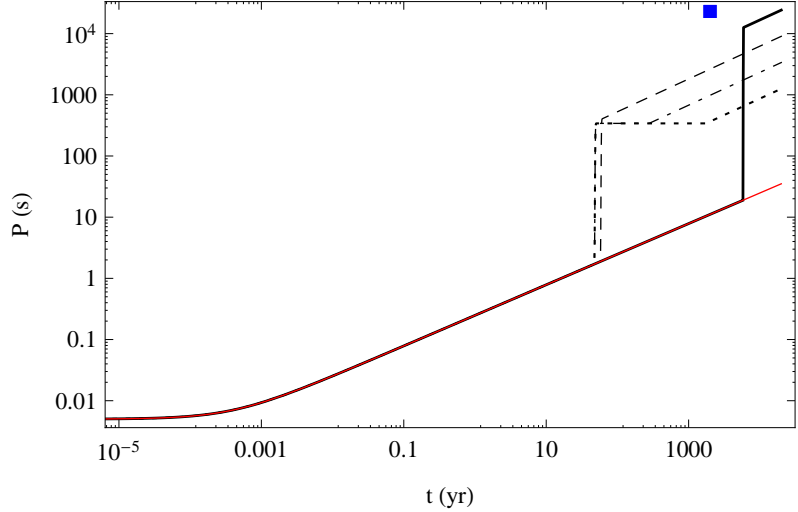}
\caption{Rotational evolution of magnetars for different masses of the fallback disk. The solid, dashed, dot-dashed, and dotted lines are for an initial disk mass of $10^{-5} \,\rm M_{\odot}$, $10^{-4} \,\rm M_{\odot}$, $10^{-3} \,\rm M_{\odot}$, $10^{-2} \,\rm M_{\odot}$, respectively. The red solid line is for an initial disk mass of $10^{-6} \,\rm M_{\odot}$. The different lines are coincide with each other at the early stage. The blue square is for the superslow magnetar in RCW 103. It is shown for comparison use only. Dedicated calculation for this source is shown in the next section.}
\label{gP_mass}
\end{figure}

The magnetic dipole field of magnetars can be in the range $10^{14}\,\rm G - 10^{15} \,\rm G$. The birth period of magnetars were thought to be much smaller than that of normal pulsars, about several milliseconds (Duncan \& Thompson 1992; Vink \& Kuiper 2006). The passive disk around magnetar 4U 0142$+$61 has a mass of $3\times 10^{-5} \,\rm M_{\odot}$, and the disk mass is not expected to vary significantly after its birth (Wang et al. 2006).  Therefore, a magnetic field of $10^{15} \,\rm G$, an initial rotational period of $5\,\rm ms$, and an initial fallback disk mass of $10^{-5} \,\rm M_{\odot}$ are chosen as typical parameters of a magnetar. Its rotational evolution is shown  in figure \ref{gP_mass}. The calculation is stopped at $2\times 10^4 \,\rm yr$, which may be the typical active life time of a fallback disk (Menou et al. 2001; Li 2007).  The magnetar will firstly be spun-down by the magnetic dipole radiation. 
In the magnetic dipole braking domain, $P^2-P_0^2 \propto t$ (eq. (5.18) in Lyne \& Grahm-Smith 2012; Tong 2016), where $P_0$ is the initial rotational period, and $P$ is the period at age $t$. During the early stage when $t$ is small, $P$ is approximately $P_0$, i.e. the rotational period is almost constant. Later when $t$ is large or equivalently $P_0$ is much smaller than the $P$ at $t$, the rotational period will increase with time as $\propto t^{1/2}$. The numerical calculation also confirms this point. 

When the magnetar has been spun-down significantly, it will enter the propeller phase. During the propeller phase, the rotational period increase with time very quickly. After the propeller phase, the magnetar tends to be in rotational equilibrium with the fallback disk. According to eq. (\ref{propeller_torque}), the equilibrium angular velocity is  $\Omega = \Omega_{\rm K} (R_{\rm m})$ (Fu \& Li 2013). In terms of rotational period: 
\begin{equation}\label{Peq}
P_{\rm eq} =2\pi \left( \frac{R_{\rm m}^3}{G M}  \right)^{1/2} =915 \, B_{15}^{6/7} \dot{M}_{\rm acc, 17}^{-3/7} \,\rm s ,
\end{equation}
where $B_{15}$ is the magnetic dipole field in units of $10^{15} \,\rm G$, and $\dot{M}_{\rm acc, 17}$ is the mass accretion rate in units of $10^{17} \,\rm g \,s^{-1}$. 
For accretion from a fallback disk, the mass accretion rate will decrease with time: $\dot{M}_{\rm acc} \propto t^{-\alpha}$. Therefore, the equilibrium period will increase with time: $P_{\rm eq} \propto t^{3\alpha/7}$. 
From figure \ref{gP_mass} it can be seen that: when the disk mass is larger, the final equilibrium period will be smaller due to a higher mass accretion rate. For a disk mass of $10^{-2} \,\rm M_{\odot}$, when the magnetar enters rotational equilibrium with the fallback disk, the mass accretion rate is still in the constant phase (see eq.(\ref{Mdotacc})). Therefore, there will be a plateau in the rotational period as a function of time. For a disk mass of $10^{-6} \,\rm M_{\odot}$, the mass accretion rate will be so low that the disk will never enter the light cylinder. The magnegtar is unaffected by the fallback disk and will  be a normal magnetar. This may correspond to the majority of  magnetars, i.e. they have no or negligible fallback accretion. 

\begin{figure}[htbp]
\centering
\includegraphics{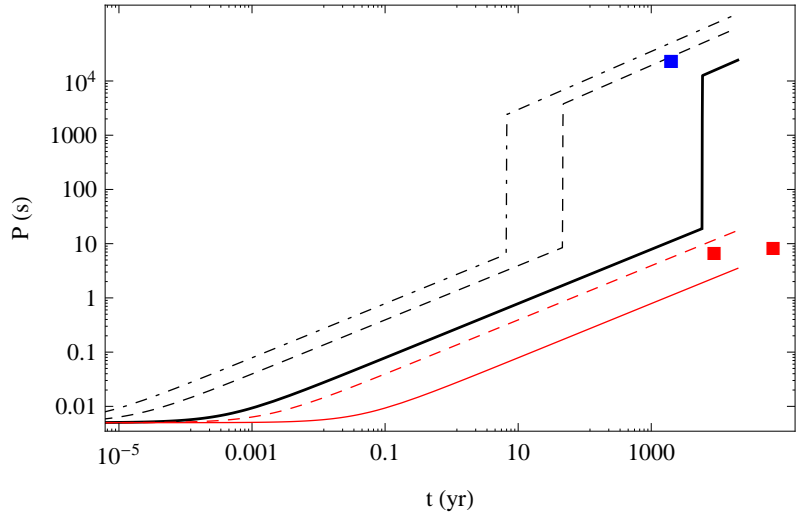}
\caption{Rotational evolution of magnetars in the presence of a fallback disk for different magnetic dipole field. The black solid, dashed, and dot-dashed lines are for magnetic dipole field of $10^{15} \,\rm G$, $5\times 10^{15} \,\rm G$, $10^{16} \,\rm G$, respectively. The red solid, and dashed lines are for magnetic dipole field of $10^{14} \,\rm G$, $5\times 10^{14} \,\rm G$, respectively. The two red squares are the observations of magnetars 4U 0142$+$61 (right) and 1E 2259$+$586 (left), respectively. }
\label{gP_B}
\end{figure}

The rotational evolution of magnetars for different magnetic dipole field is shown in figure \ref{gP_B}. 
The initial rotational period is chosen as $5\,\rm ms$, the initial disk mass $10^{-5} \,\rm M_{\odot}$. 
From figure \ref{gP_B}, when the magnetic dipole field is large, the magnetar can be significantly spun-down during the dipole braking stage. It will enter the propeller stage and be spun-down very quickly later.  This may correspond to the super-slow magnetar in RCW 103. 
When the surface dipole field is lower than several times of $10^{14} \,\rm G$, the magnetar can not be significantly  spun-down and will always be in the ejector phase. In this case, the fallback disk may be seen when illuminated by the central magnetar. Observationally, two magnetars 4U 0142$+$61 and 1E 2259$+$586 may have fallback disks (Wang et al. 2006; Kaplan et al. 2009). Timing observations of these two magnetars showed that their surface dipole field are both around or smaller than $10^{14} \,\rm G$, depending on the assumed braking mechanism (Tong et al. 2013; Olausen \& Kaspi 2014). The two observed magnetar/fallback disk systems are consistent with the calculations here\footnote{The discrepancy between the supernova remnant age and the characteristic age of magnetar 1E 2259$+$586 may involve additional processes (Rogers \& Safi-Harb 2016).}. Therefore, 
when the disk mass is substantial, the system can be seen either as a normal magnetar/fallback disk system or a super-slow magnetar, depending on the magnetic dipole field. The three different kinds of magnetars can be understand together in the magnetar/fallback disk system. 

\begin{figure}[htbp]
\centering
\includegraphics{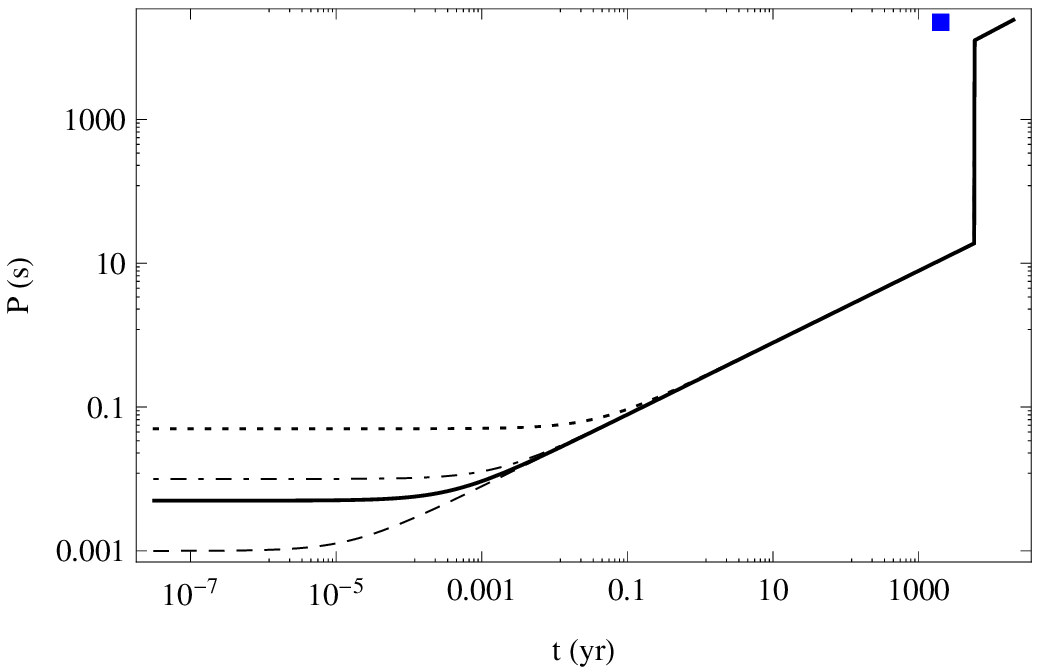}
\caption{Rotational evolution of magnetars in the presence of a fallback disk for different initial rotational period. The dashed, solid, dot-dashed, and dotted lines are for initial rotational period of $1\,\rm ms$, $5\,\rm ms$, $10\,\rm ms$, $50\,\rm ms$, respectively. }
\label{gP_P}
\end{figure}

Previously, a long initial period is required when considering the fallback disk accreting magnetar model (De Luca et al. 2006; Li 2007). The rotational evolution of magnetars for different initial rotational period is shown in figure \ref{gP_P}. The magnetic dipole field is chosen as $10^{15} \,\rm G$, and fallback disk mass $10^{-5} \,\rm M_{\odot}$. Figure \ref{gP_P} shows that the information of initial rotational period are lost even during the magnetic dipole braking stage, and the later evolution are insensitive to the choice of initial rotational period. 

\subsection{Calculation for the super-slow magnetar in RCW 103}

\begin{figure}[htbp]
\centering
\includegraphics{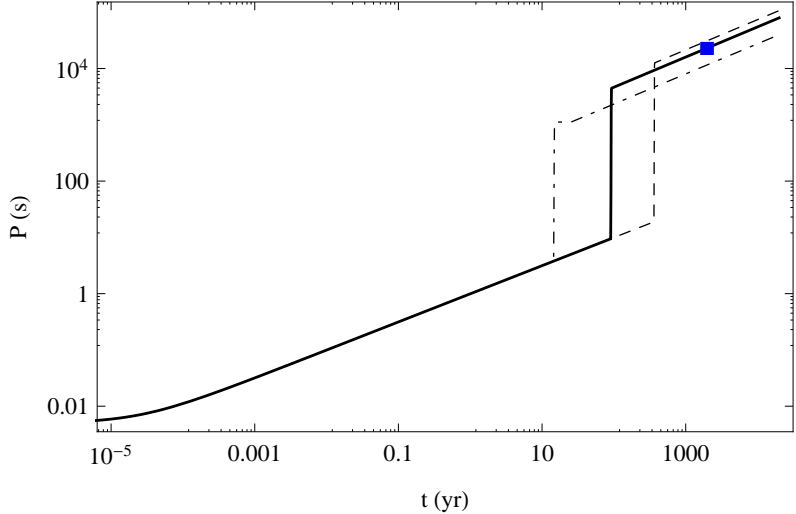}
\caption{Rotational evolution of a high magnetic field magnetar ($B=4\times 10^{15} \,\rm G$) in the presence of a fallback disk. The blue square is the superslow magnetar in RCW 103. The dashed, solid, and dot-dashed lines are for an initial fallback disk mass of $0.5\times 10^{-5} \,\rm M_{\odot}$, $10^{-5} \,\rm M_{\odot}$, and $5\times 10^{-5} \,\rm M_{\odot}$, respectively.}
\label{gP_rcw103}
\end{figure}

The CCO inside the supernova remnant RCW 103 is identified as a magnetar (D'Ai et al. 2016; Rea et al. 2016). The supernova remnant age is about $2 \,\rm kyr$ (De Luca et al. 2006). The $2.4\times 10^{4} \,\rm s$ period is probably the magnetar's rotational period (De Luca et al. 2006; Esposito et al. 2011; D'Ai et al. 2016; Rea et al. 2016). From the above calculations, especially figure \ref{gP_B}, this super-slow magnetar may have a very high magnetic dipole field. Figure \ref{gP_rcw103} shows the rotational evolution of a high magnetic field magnetar with different fallback disk masses. The magnetic dipole field is chosen as $4\times 10^{15} \,\rm G$, and an initial rotational period of $5\,\rm ms$. For an initial fallback disk mass around $10^{-5} \,\rm M_{\odot}$, the magnetar can be significantly spun-down in less than $2 \,\rm kyr$. 

From eq.(\ref{Peq}), if the fallback disk around the super-slow magnetar is still active, then a large period derivative is expected: $\dot{P} = 3\alpha/7 \times P/t \approx 2\times 10^{-7}$. However, the observational upper limit on the period derivative is $|\dot{P}|<1.6\times 10^{-9}$ (Esposito et al. 2011). This may because the fallback disk have become neutralised and inactive. The typical life time of a fallback disk is about several thousands of years (Menou et al. 2001). It is possible that the fallback disk around the super-slow magnetar in RCW 103 has already become inactive at an age of $2\,\rm kyr$. The magnetar is now spun-down by magnetic dipole braking. The expected period derivative is about $10^{-12}$.  It is consistent with present upper limits. It also means that the persistent X-ray luminosity and outburst of the super-slow magnetar originate from the magnetic energy. The outburst of the super-slow magnetar indeed resemble those of typical magnetars (Rea \& Esposito 2011; D'Ai et al. 2016; Rea et al. 2016). Though the disk activity has faded away, the magnetic dipole field is not expected to decay significantly in $2\,\rm kyr$ (Vigano et al. 2013). At present, the magnetar SGR 1806$-$20 has the highest characteristic magnetic field, about $2.5\times 10^{15} \,\rm G$ (Tong 2013; Olausen \& Kaspi 2014). The expected magnetic dipole field of the super-slow magnetar should be around $4\times 10^{15} \,\rm G$. A little bit higher than that of magnetar SGR 1806$-$20. In the above scenario, the X-ray emissions of the super-slow magnetar originate from the magnetic energy. Therefore, previous spectral models for magnetars may also be applied to this source (Weng et al. 2015). If future spectra modeling can give some information about the magnetic dipole field of this super-slow magnetar, the model presented here can be tested. 

\subsection{Accretion induced magnetic field decay}

After the supernova explosion, there may be a hypercritical accretion phase. The initial magnetic field may be buried by the accreted matter. This may account for the existence of normal radio pulsars and radio quiet pulsars (for example CCOs)  (Geppert et al. 1999; Geppert 2009; Vigano \& Pons 2012). However, the magnetar's high magnetic field and rapid rotation can prevent accretion onto the central neutron star, even when the accretion is hypercritical (Geppert 2009). For neutron stars in X-ray binaries, an empirical relation for the accretion induced magnetic field decay is: $B = B_0/(1+\Delta M/10^{-4} \,\rm M_{\odot})$ (Shibazaki et al. 1989; Zhang \& Kojima 2006), where $B$ and $B_0$ are the reduced magnetic field after accretion and the initial magnetic field, respectively, and $\Delta M$ is the accreted mass. 
For a typical disk mass of $10^{-5} \,\rm M_{\odot}$,  the disk mass itself is relatively small. Furthermore, from figure \ref{gP_mass}, at the ejector phase and propeller phase, most of the accreted matter can not fall onto the neutron star. Similarly, for a disk mass of $10^{-2} \,\rm M_{\odot}$, most of the disk mass can not fall onto the neutron star. 
According to this empirical relation, the evolution of the magnetar magnetic field is negligible in the case of a fallback disk. Therefore, the assumption of a constant magnetic field for a magnetar in the presence of a fallback disk is reasonable. 

\subsection{Different propeller torques}

In the above calculations, the accretion torque is adopted from Menou et al. (1999) and Chatterjee et al. (2000), see eq. (\ref{propeller_torque}). It is an educated guess containing the spin-up torque and the propeller torque. When the accretion torque vanishes, it corresponds to accretion equilibrium. A more general form of accretion torque can be written as:
\begin{equation}
N_{\rm prop} \propto \dot{M}_{\rm acc} R_{\rm m}^2 \Omega_{\rm K}(R_{\rm m}) \left(  1-  \left(\frac{\Omega}{\Omega_{\rm K}(R_{\rm m})} \right)^{\chi} \right), 
\end{equation}
where $\chi$ is a free parameter. Different values of $\chi$ correspond to different propeller torques. When $\chi =1$, it corresponds to the torque of Menou et al. (1999) and Chatterjee et al. (2000). When $\chi =2$, it corresponds to the torque of Benli \& Ertan (2016). When $\chi =1/2$  it corresponds to the torque of Liu et al. (2014). $\chi$ can also have other values (Francischelli \& Wijers 2002; Ghosh 1995; Ertan et al. 2007). Irrespective of the exact value of $\chi$, in accretion equilibrium, the rotational period is always given by the equilibrium period, see eq.(\ref{Peq}). The predicted period derivative is also the same, if the fallback disk is still active. According to eq.(\ref{Peq}), the super-slow magnetar in RCW 103 should be a high magnetic field magnetar with an active fallback disk in the past. A high magnetic field ensures a large equilibrium period. The fallback disk has now becomes inactive. This will result in a small period derivative. This conclusion is rather independent of the exact form of accretion torques. It only requires that the magnetospheric radius is equal to the corotation radius during accretion equilibrium (Lai 2014).

The central neutron star is spun-down quickly during the propeller phase before entering accretion equilibrium. The above calculations are for the case of $\chi=1$. In the propeller phase, $\Omega/\Omega_{\rm K}(R_{\rm m}) >1$. Therefore, for $\chi>1$ (e.g. Benli \& Ertan 2016), the propeller spin-down is more efficient than that of $\chi=1$ (Menou et al. 1999). The neutron star will enter into accretion equilibrium earlier. For $\chi<1$ (e.g. Liu et al. 2014), the propeller spin-down is less efficient than that of $\chi=1$. The neutron star will enter into accretion equilibrium a little bit later. Numerical calculations show that there is only marginal difference in the propeller phase for the three different torques, consistent with the above qualitative analysis. During the ejector phase and accretion equilibrium phase, the results are the same for the three different torques (Menou et al. 1999; Liu et al. 2014; Benli \& Ertan 2016). 

\section{Discussion and conclusion}

For a normal neutron star/fallback disk system, the neutron star will first enter the accretor/propeller phase. Then it will enter the ejector phase or acquire rotational equilibrium with the fallback disk (Chatterjee et al. 2000; Yan et al. 2012; Fu \& Li 2013; Benli \& Ertan 2016). From the above calculations for a magnetar/fallback disk system, the magnetar will be first in the ejector phase. After it has been significantly spun-down, it may enter the propeller phase and then acquire rotational equilibrium with the fallback disk. This is the difference between a magnetar/fallback disk system and a normal neutron star/fallback disk system. 

Ultra-luminous X-ray pulsar is a neutron star in a binary system whose X-ray luminosity can be as high as $10^{40} \,\rm erg \, s^{-1}$ (Bachetti et al. 2014). The super-Eddington luminosity may be due to the presence of magnetar strength magnetic field. Ultra-luminous X-ray pulsar may be an accreting magnetar in a binary system (Tong et al. 2015). 
Super-slow X-ray pulsars are neutron stars in binary systems having rotational period longer than $10^{3} \,\rm s$. If they rotate at the equilibrium period for a given accretion rate, their dipole field can be as high as $10^{15} \,\rm G$ (as can be seen in eq.(\ref{Peq})). Therefore, the accreting magnetar scenario is also employed to explain the super-slow X-ray pulsars (Popov \& Turolla 2012; Wang 2013).

According to the above calculations, the fallback disk can affect the rotational evolution of the magnetars only when the disk mass is substantial and the magnetic dipole field is very high. Otherwise, the fallback disk is either negligible or can only be seen as a passive fallback disk. Only two magnetars has possible fallback disks observed. This may due to the magnetar's relatively low magnetospheric activities in the optical/infrared band and high X-ray luminosity (Wang et al. 2006). At the same time, it can not be excluded that magnetars are formed through peculiar channels, e.g., binary origin (Clark et al. 2014; Popov 2015) or descendent from Thorne-Zytkow object (Liu et al. 2015). 

In the calculations, typical parameters of magnetars and fallback disks are chosen according to the current understandings. In the future, if the dipole field of the super-slow magnetar in RCW 103 is found to be inconsistent with the calculations here, it may imply that our current understanding of magnetars and/or fallback disks is still not perfect. 

In conclusion, for the rotational evolution of magnetars in the presence of a fallback disk, three different kinds of magnetars can be formed in a unified way:
\begin{enumerate}
\item When the disk mass is lower than $10^{-6} \,\rm M_{\odot}$, the magnetar is aways in the ejector phase and will be a normal magnetar. This may correspond to the majority of magnetars. 

\item When the disk mass is substantial, e.g., about $10^{-5} \,\rm M_{\odot}$, and the magnetar dipole field is relatively low, e.g. about or smaller than $10^{14} \,\rm G$, the magnetar will also be a normal magnetar. The fallback disk may be seen when illuminated by the central magnetar. This may correspond to the two observed magnetars with possible fallback disks (Wang et al. 2006; Kaplan et al. 2009). 

\item When the disk mass is substantial and the magnetic dipole field is very high, e.g., around $4\times 10^{15} \,\rm G$, the magnetar will be significantly spun-down even when it is still young. This may correspond to the super-slow magnetar in RCW 103 (De Luca et al. 2006; Esposito et al. 2011; D'Ai et al. 2016; Rea et al. 2016). Future information about its magnetic dipole field can test this model. 

\end{enumerate}
In addition, the magnetar's initial rotational period is not important.

\section*{Acknowledgments}
H.Tong is supported by 973 Program (2015CB857100), Qing Cu Hui of CAS, and NSFC (U1531137).

\end{document}